\documentclass[prb,
twocolumn,showpacs,
floatfix
]{revtex4}

\usepackage{graphicx}
\usepackage{bm}
\newcommand{\bq}{\begin{equation}}
\newcommand{\eq}{\end{equation}}

\newcommand{\ev}{\end{verse}}

\newcommand{\bv}{\begin{verse}}
\newcommand{\bfk}{{\bf k}}

\newcommand{\fm}{(-{\partial f\over\partial\epsilon})}

\begin{document}

\title{Hall effect in the marginal Fermi liquid regime of high-$T_c$
superconductors}

\author{Elihu Abrahams}
\affiliation{Center for Materials Theory, Serin Physics
Laboratory, Rutgers University, Piscataway, NJ
08854-8019}
\author{C.M. Varma} \affiliation{Bell Laboratories,
Lucent Technologies Murray  Hill, NJ 07974}

\date{\today}

\begin{abstract}
The detailed derivation of a theory for transport in quasi-two-dimensional metals,
with small-angle elastic scattering and angle-independent inelastic scattering
is presented. The transport equation is solved for 
a model Fermi surface representing a typical cuprate superconductor. Using the
small-angle elastic and the inelastic scattering rates deduced from angle-resolved
photoemission experiments, good quantitiative agreement with the observed anomalous
temperature dependence of the Hall angle in {\it optimally} doped cuprates is obtained,
while the resistivity remains linear in temperature. The theory is also extended to the
frequency-dependent complex Hall angle.

\end{abstract}

\maketitle
\section{Introduction}
Soon after the discovery of high-temperature
superconductivity in the cuprate compounds, it was found that all 
the transport properties for compositions near those for
the maximum $T_c$ have anomalous temperature and/or frequency dependence, 
in contrast to what is expected for Landau
Fermi liquids. Most of the anomalies, in
resistivity and optical conductivity, in the frequency dependence
of Raman intensity, in the tunneling spectra, and in the temperature
dependence of the copper nuclear relaxation could be understood
by the ``Marginal Fermi Liquid" (MFL) phenomenology.\cite{mfl} Within
MFL, scale-invariant fluctuations in both in the charge and  magnetic
sectors are assumed to have
the form,
\begin{eqnarray}
{\rm Im}\chi({\bf q},\omega,T) &= & -N_0 (\omega/T),
\;\;\;\;\;\;  \omega \ll T \nonumber\\
& = & -N_0 ({\rm sgn}\omega), \;\;\;\;\,\,
T\ll\omega\ll\omega_c,
\end{eqnarray}
characteristic of a quantum critical point. $\omega$ is the frequency and
${\bf{q}}$ the momentum of the fluctuations, $N_0$ is the density of
energy states per unit volume and $\omega_c$ is a
high-frequency cutoff. At long wavelengths this form is observed directly in
Raman scattering\cite{Slakey}. A principal prediction\cite{mfl} from this
hypothesis is that at low energies ($\omega \ll T$), the inelastic part
of the single-particle relaxation rate has the MFL form $\lambda T$ with
coefficient $\lambda$ having negligible momentum dependence either along
or perpendicular to the Fermi surface. This form has been confirmed in
angle-resolved photoemission (ARPES) experiments\cite{arpes} and
leads directly to the observed linear temperature dependence of the
resistivity.

However, the temperature dependence of anomalies in the normal state
Hall effect and magnetoresistance could not be understood by the MFL
hypothesis alone. For example, for a situation near optimal doping,
where the resistivity is linear in temperature, the expectation is
that the cotangent  of the Hall angle ($\sigma^{xx}/\sigma^{xy}$)
should also be linear. Experiment shows it to be more nearly
quadratic. This left open the possibility that essential new physics near
optimum doping may not be captured by the MFL scenario. In this paper we
show that the temperature dependence of the Hall
angle may be understood quantitatively by a proper application of transport theory
using the measured single-particle relaxation rates.  

The single-particle self energy
$\Sigma({\bf k},\omega,  T)$ is measurable in angle-resolved
photoemission (ARPES) experiments.\cite{arpes,pnas} As stated above,
within MFL, the prediction is that the inelastic part of
$\Sigma$ is 
${\bf k}$-independent and of the MFL form,\cite{mfl}
proportional to $\omega$ for $\omega \gg T$ and to $T$ for $T \gg
\omega$. ARPES experiments \cite{arpes} do find the
inelastic part $\Sigma$ of this form but find in
addition an elastic part which varies in magnitude around the
Fermi surface. Thus, as explained in detail earlier,\cite{pnas} on the
Fermi surface ($\omega = 0$), the experimentally-measured self energy
consists of the MFL part $\lambda T$ and an anisotropic
$T$-independent elastic part:
\begin{equation}
{\rm Im}\: \Sigma ({\vec k},T)\  = \lambda T + \gamma(\hat k),
\end{equation}

According to experiment,
$\gamma(\hat k)$ 
increases by about a factor 4 to 5 going from the $(\pi,\pi)$ to the
$(\pi,0)$ direction along the Fermi surface.\cite{pnas} A crucial point
is that even at its minimum value, it is more than an order of magnitude 
larger than the transport scattering rate due to impurities
obtained by extrapolating the normal state resistivity to $T=0$.
As argued in Ref.\ 4, such  behavior arises if $\gamma(\hat k)$
comes from small-angle impurity scattering. We discuss this point further
in Sec.\ VII. 

In a previous communication,\cite{prl} we described, for
high-$T_c$ superconductors, how Eq.\ (1) can be used in a Boltzmann
equation analysis to account for observed 
anomalies\cite{ong,exp,hwang,ando,raffy,drew} in the Hall effect.   We
performed a calculation (page 4655 of Ref.\ 1) using a
simple Fermi surface in order to give an example of how a new
contribution \cite{corr} could dominate the conventional result
and thus account for experimental observations. However it was
pointed out to us by V. Yakovenko \cite{vy} that we erred in the
form we chose to parameterize $\gamma(\hat k)$.\cite{err} The purposes of 
this paper are to give a more
complete solution of the Boltzmann equation, correct the above error,
and to present a model calculation which illustrates how, with a proper
parameterization and a reasonable choice of parameters, the presence of
the anisotropic impurity scattering together with other anisotropies can
account for the observed behavior of the Hall
angle.\cite{ong,exp,hwang,ando,raffy,drew}
Further, we generalize our results to the complex Hall
angle at finite frequencies which has been recently measured.\cite{matt}

A point about the experimental results needs stressing. In many
presentations of the data, for example Refs.\ 9,10, single temperature
non-integer power-law fits to the Hall number or the Hall angle have been made,
with varying success, while integer power law fits \cite{ong} do not
usually faithfully represent the data. There is no physical reason to
expect integer power law behavior and our
analysis shows quite generally that the solution of the Boltzmann equation can
give a sum of contributions with different temperature dependences. In
particular, it is usually found\cite{ando,raffy} that a temperature power law
with a power less than 2 can fit the data for the cotangent of the Hall angle. As
we show here, this is indistinguishable from the ratio of terms each of
which is a sum of contributions with different temperature dependences.

Our results have two significant conclusions. First, anisotropies, especially
in the scattering rate, introduce corrections
to the calculation of magnetotransport properties so that conventional
results such as $R_H = (nec)^{-1}$, or $\tan\theta_H = \omega_c\tau_{tr}$
are not in general valid.\cite{hlu1} Second, the fact that the  Hall effect
in the cuprates can be understood within transport
theory, with only the measured single-particle relaxation
rate, the measured  resistivity and Fermi surface quantities as
inputs implies that no new physics is involved other than what leads to
the behavior of those quantities. We conclude that the MFL, which
gives the experimentally observed temperature dependence of the
relaxation rate, is sufficient to account for all the the anomalous
magnetotransport near optimal doping.
This point of view is reinforced by recent
frequency-dependent complex Hall effect measurements\cite{matt} whose
results are consistent with the analysis presented here, as discussed in
Sec.\ VI.

\section{Transport Equation}
At long wavelengths and low
frequencies the derivation of Boltzmann equation relies only on
conservation laws and is valid whether the system is a
Fermi liquid or not.
Therefore we start with the linearized Boltzmann equation for the
deviation
$g({\bf{k}},t)$ of
the momentum distribution function from its equilibrium value
$f({\bf{k}})$
in the presence of uniform
static electric and magnetic fields,
\bq
\frac{\partial g_{\bfk}}{\partial t} + e{\bf E}\cdot{\bf
v}_{\bfk}\;\frac{\partial f}{\partial\epsilon_{\bfk}} +
\frac{e}{\hbar c}({\bf v}_{\bfk}\times{\bf B})\cdot\nabla_{\bfk}\,
g_{\bfk} = C_{\bfk}
\eq
Here $C({\bf{k}})$ is the collision operator.
The solution to this equation for the stationary case ($\omega = 0$) is
given by
\cite{KSV}
\begin{equation}
g({\bf k})=e\hbar\sum _{\bf k'}\  A^{-1}_{{\bfk},{\bfk}'}\left[{\bf E}
\cdot {\bf v}_{{\bf k}'}\left(
-\frac{\partial f}{\partial
\epsilon _{{\bf k}'}}
\right)\right],
\end{equation}
where
\begin{equation}
A_{{\bf k},{\bf k}'}=\hbar\left[{1\over
\tau({\bf k}) }
\,  +{e\over \hbar c}\:
{\bf v}_{\bf
k}\times {\bf B}\cdot \nabla _{\bf k}\right]\delta_
{{\bf
k},{\bf k}'}-C_{{\bf k},{\bf k}'}.
\end{equation}
Here $C_{{\bf k},{\bf k}'}$ is the ``scattering-in''
term
in the collision operator for the Boltzmann  equation
and
$ \hbar/\tau({\bf k})
=\sum _{{\bf k}'}C_{{\bf k},{\bf k}'}$,
is
the ``scattering-out'' term equal to the
single-particle
relaxation rate. It is evident that the
distribution
$g({\bf k})$ is not only determined by
the
energy of the state ${\bf k}$
but also by the anisotropy
of the scattering. The
distribution is depleted in
directions of large net scattering
and augmented in
directions of small net scattering. 

We calculate the conductivities 
using the single-particle scattering rate
of Eq.\
(2). The
conductivity tensor is
\bq
\sigma ^{\mu \nu}=\frac{e^{2}\hbar}{\Omega}\sum _{{\bf
k},
{\bf k}'}\:
v _{\mu ,{\bf k}}\  A_{\bf k,\bf k'
}^{-1} \:
v_{\nu ,{\bf k}'}\left(-{\partial f\over
\partial
\epsilon _{\bf k '}} \right),
\eq
where $\Omega$ is the sample (taken to be a plane) area.
We expand
$A^{-1}$ from Eq.\ (4) in powers of
$\bf B$: \cite{KSV}
\begin{eqnarray}
A^{-1}={\cal T} &-&(e/c){\cal T} ({\bf
v}\times {\bf B\cdot \nabla })\,
{\cal T} \nonumber \\
&+&(e/c)^2{\cal T}
({\bf v\times B\cdot \nabla})\,{\cal
T}\,({\bf v\times B\cdot
\nabla })\,{\cal T},
\end{eqnarray}
where (from Eq.\ (2.12) of Ref.\ 17)
\bq
{\cal T}_{\bfk,\bfk'} = {1\over\hbar}[\tau_{\bfk}\;
\delta_{\bfk,\bfk'}
+
\sum_{\bfk''}
\tau_{\bfk}~C_{\bfk,\bfk''}\;
{\cal T}_{\bfk'',\bfk'}]
\eq
The first term in $A^{-1}$ gives the longitudinal
conductivity,
the second the Hall conductivity
$\sigma^{xy}$
and the
magnetoconductivity may be
calculated from the third term.

As discussed above, the $\tau_{\bfk}$ is the MFL scattering plus
the angle dependent impurity piece:
\bq
1/\tau_{\bfk} = 1/\tau_M +
1/\tau_i\,(\hat k).
\eq
Our approach to solving Eq. (7) is to take
advantage of the properties of $1/\tau_M$ and $1/\tau_i$.
The kernel $C$ in Eq.\ (8) 
comprises the vertex corrections for the various scattering
mechanisms. Within MFL, the $1/\tau_M$ is an $s$-wave scattering
process so the MFL interaction does not appear in $C$. That leaves
\bq
C_{{\bf k},{\bf k}'}=  2 \pi\delta (\epsilon_{\bf k} -
\epsilon_{{\bf k}'})|u_{i}(\theta ,\theta^{\prime})|^2,
\eq
where,
e.g., $\theta$ is the angle of ${\hat k}$ along the Fermi surface.
We have argued \cite{prl,pnas} that the impurity scattering matrix
element $u_i(\theta,\theta')$ involves scattering through small
angles only. This property allows an expansion of the RHS of Eq.\
(8) in powers of a ``small angle scattering parameter", $\theta_c$
which can itself be a function of $\theta$. We illustrate such an
expansion for the calculation of $1/\tau_i$:
\begin{eqnarray}
{1\over
\tau_i}(\theta) &=& {1\over
\hbar}\sum_{\bfk'}|u_{i}(\theta,\theta^{\prime})|^2\delta(\epsilon'-
\epsilon) \nonumber \\
&\approx& \theta_c\,U(\theta,\theta)N(\theta) + {\theta_c^3\over
24}\left[{d^2\over
d\theta'^2}U(\theta,\theta')N(\theta')\right]_{\theta'=\theta}.
\end{eqnarray} 
Here we have used
\bq
\sum_\bfk = {1\over2\pi}\int d\epsilon\int 
N(\theta)d\theta, ~~~
N(\theta) = {\Omega\over 
2\pi\hbar}\,{d k_t/d\theta\over v(\theta)}.
\eq
$N(\theta)$ is the density of states per unit energy
at 
the Fermi surface at angle $\theta$ and $v(\theta)$ is the 
Fermi
velocity. $dk_t$ is an infinitesimal taken tangent to the Fermi
surface at the angle $\theta$. We have abbreviated 
$|u_i(\theta,\theta')|^2$ by
$U(\theta,\theta')$. We shall describe below how experiments
show that $\theta_c$ is indeed sufficiently small to justify such an
expansion.

\section{Longitudinal Conductivity}
We begin by determining the longitudinal conductivity within the
model. We first give the solution for a general Fermi surface  and
then specialize to a particular example to illustrate the results.

From Eqs. (5,6), we have
\bq
\sigma^{xx} = {e^2\hbar\over\Omega}\sum_{{\bf k}{\bf k}'}v_{{\bf
k}}^x~{\cal
T}_{{\bf k},{\bf k}'}~
v_{{\bf k}'}^x(-{\partial f\over \partial\epsilon'})
\eq
Define the vector {\bf L} as
\bq
{\bf L_\bfk} = \sum_{\bfk'}{\cal
T}(\bfk,\bfk')(-\partial
f/\partial\epsilon'){\bf v_{\bfk'}},
\eq
so that
\bq
\sigma^{xx} = {e^2\hbar\over\Omega}\sum_{\bfk} v_{\bfk}^x L_{\bfk}^x.
\eq
Using Eq.\ (8) in Eq.\ (14), we find the integral equation for ${\bf
L}$:
\bq
{\bf L_\bfk} = {\tau_\bfk\over\hbar}\left[\fm v_\bfk^x \,+ 
\sum_{\bfk''}
C_{\bfk,\bfk''}\,{\bf L_{\bfk''}}\right].
\eq
As discussed earlier, $C$ contains only the small angle 
scattering so
that
\bq
{\bf L_\bfk} = {\tau_\bfk\over\hbar}\left[\fm {\bf v_\bfk} \,+
\int d\theta' N(\theta') 
U(\theta,\theta')\,{\bf L_{\bfk'}}\right],
\eq
which, since ${\bf L}$ is restricted to the Fermi surface, can be
rewritten as
\begin{eqnarray}
{\hbar\over\tau(\theta)}{\bf L}(\theta) &=& \fm{\bf v}(\theta) +
{\hbar\over\tau_i(\theta)}{\bf L}(\theta) \nonumber \\ &+& \int d\theta'
N(\theta')  U(\theta,\theta')[{\bf L}(\theta')
- {\bf L}(\theta)],
\end{eqnarray}
where $1/\tau_i = 1/\tau-1/\tau_M$ is given in Eq.\ (11). Given the
small-angle scattering restriction on $U(\theta,\theta')$, we expand 
the difference ${\bf L}(\theta')-{\bf L}(\theta)$ and find the
differential equation (primes indicate derivatives with respect to
$\theta$)
\bq
{\bf M}(\theta)= {\bf
v}(\theta) + 
u(\theta){\bf M}''(\theta) + 2 u'(\theta) {\bf M}'(\theta),
\eq
where 
\bq
{\bf M}(\theta) = \frac{\hbar}{\tau_M(-\partial f/\partial \epsilon)}{\bf
L}(\theta),
\eq
and
\begin{eqnarray}
u(\theta) &=&
(\theta_c^3/24)(\tau_M/\hbar)N(\theta)U(\theta,\theta),\nonumber \\ u'(\theta)
&=& 
(\theta_c^3/24)(\tau_M/\hbar)[\frac{d}{d\theta'}N(\theta')U(\theta,\theta')]
_
{\theta'\rightarrow\theta}.
\end{eqnarray}

This differential equation can be solved for the components of ${\bf M}$,
hence ${\bf L}$, once the $\theta$-dependences of ${\bf v},\,u$ and $u'$
are known. This equation, which is basic to the evaluation of both the
longitudinal and the Hall conductivities is completely equivalent 
to Eq.\ (10) of Ref.\ 5. Although $u$ contains a small parameter an iterative
solution of Eq.\ (19), as in our previous work,\cite{prl} is not
valid. We solve Eq. (19) exactly in this paper.

As we discussed previously,\cite{prl,err} the details of the 
anisotropies of the quantities entering the transport coefficients 
determine the magnitudes of the various contributions, in particular to 
$\sigma^{xy}$ in the presence of a magnetic field. The shape of the 
Fermi surface is especially important since it determines the size of 
the contribution to $\sigma^{xy}$  from the isotropic marginal Fermi 
liquid scattering rate $1/\tau_M$. In what follows, we take the 
anisotropy of the impurity scattering from ARPES data and we assume a 
simple form for the Fermi surface velocity.

\begin{figure}
\includegraphics[width=6cm]{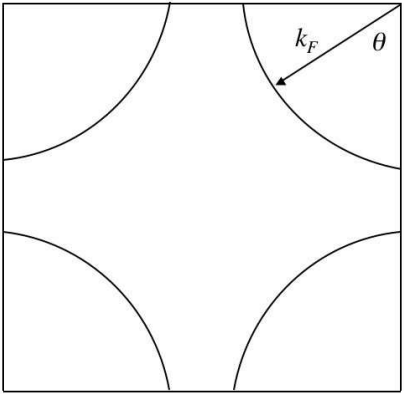}
\caption{Schematic Fermi surface.}
\label{fig1}
\end{figure}

A schematic of the Fermi surface is shown in Fig.\ 1, where the angular 
variable $\theta$ is shown.  A general Fermi surface respecting the
square
symmetry of the CuO$_2$ layers
must have Fermi velocities in the first quadrant of the form
\begin{eqnarray}
v_x &=& \sum_{n} v_n \sin[(2n+1)\theta] \nonumber  \\
v_y &=& \sum_{n} v_n (-1)^n \cos[(2n+1)\theta].
\end{eqnarray}
For our calculations we take a form for ${\bf v}(\theta)$ which is the
simplest extension beyond the circular Fermi surface (similar to that shown
in Fig.\ 1) consistent with Eq.\ (22). In the first quadrant,
\begin{eqnarray}
v_x &=& v_0(\sin\theta + \rho\sin 3\theta) \nonumber \\
v_y &=& v_0(\cos\theta -\rho\cos 3\theta) 
\end{eqnarray}

For this choice for ${\bf v}(\theta)$, the density of
states of Eq.\ (12) in all quadrants is
\bq
N(\theta) =
\frac{\Omega}{2\pi\hbar}~\frac{k_0(1-\rho)^{1/4}}{v_0(1-\rho\cos
4\theta)^{5/4}},
\eq
where $k_0$ is the value of $k_F(\theta)$ at $\theta=0$.

Thus, according to Eq.\ (15),  $\sigma^{xx}$  is given by:
\bq
\sigma^{xx} = \frac{e^2 \tau_M
(1-\rho)^{1/4}}{\pi^2
\hbar}~\frac{k_0}{v_0}\int_0^{\pi/2}d\theta\frac
{v_x(\theta)M_x(\theta)}{(1-\rho\cos
4\theta)^{5/4}}.
\eq

\section{Hall Conductivity}
For a magnetic field
perpendicular to the $xy$ plane, ${\bf v}\times {\bf
B}\cdot\nabla =
[\Omega/2\pi\hbar N(\theta)]B\cdot\partial/\partial
\theta$ and from Eqs.\ (6,7),
\bq
\sigma^{xy} = {e^3B \over 2\pi c}\sum_{\bfk,\bfk',\bfk1}
v_\bfk^x\;\frac{{\cal T}(\bfk,\bfk_1)}{
N(\theta_1)}{d{\cal T}(\bfk_1,\bfk')\over
d\theta_1} (-{\partial
f\over\partial\epsilon'})\,v_{\bfk'}^y.
\eq
We rewrite this using Eq.\ (14):
\bq
\sigma^{xy} = {e^3B\over 2\pi c}\sum_{\bfk,\bfk1}
v_\bfk^x\;\frac{{\cal T}(\bfk,\bfk_1)}{
N(\theta_1)}{d\over d\theta_1} L_{\bfk 1}^y.
\eq

We define
\bq
{\bf K_{\bfk}} = \sum_{\bfk'}{\cal
T}(\bfk,\bfk'){1\over N(\theta')}{d\over d\theta'}{\bf L_{\bfk'}},
\eq
so that
\bq
\sigma^{xy} = {e^3B\over 2\pi c}\sum_{\bfk}v_\bfk^xK_\bfk^y.
\eq
Using Eq.\ (16) in Eq.\ (31), we get an integral equation for 
${\bf K}$:
\begin{eqnarray}
{\bf K_\bfk} &=& {\tau_\bfk\over\hbar}\left\{{1\over N(\theta)}
{\bf L_{\bfk}}' +
\sum_{\bfk''}C_{\bfk,\bfk''}{\bf K_{\bfk''}}\right\} \nonumber \\
&=& 
{\tau_\bfk\over\hbar}
\left\{{{\bf L_{\bfk}}'\over N(\theta)}
 + \int d\theta'N(\theta')U(\theta,\theta'){\bf
K}(\theta')\right\}.
\end{eqnarray}

We proceed exactly as in the analysis for ${\bf L}$, Eqs.\ (16-21).
For the required $K_y$, we find the differential equation 
\bq
Z(\theta) = M_y'(\theta)(1-\rho\cos 4\theta)^{5/4} + u(\theta)Z''(\theta)
+ 2u'(\theta)Z'(\theta)
\eq
where $Z(\theta)$ determines $K_y(\theta)$ as
\bq
K_y(\theta) = 
\left(\frac{\tau_M}{\hbar}\right)^2{2\pi\over \Omega}{\hbar v_0\over k_0}
{1\over (1-\rho)^{1/4}}\fm ~Z(\theta).
\eq
and $M_y$ is obtained as the solution of Eq.\ (19). As in that case, an
iterative solution of Eq.\ (31) is not valid. 
Combining all our results into
Eq. (29), we find
\bq
\sigma^{xy} = -{e^3B\over \pi^2c}\left({\tau_M\over\hbar}\right)^2
\int_0^{\pi/2} \frac{d\theta}{(1-\rho\cos 4\theta)^{5/4}}~
v_x(\theta)Z(\theta)
\eq

When there is no small-angle scattering, $M_y=v_y$ and $Z(\theta) = -
v_0(\sin\theta -3\rho\sin 3\theta)(1-\rho\cos 4\theta)^{5/4}$. In that
case,
\bq
\sigma^{xy} \propto \int_0^{\pi/2} d\theta 
(\sin^2\theta - 3\rho^2\sin^2\theta)
\eq
which
vanishes when
$\rho = 1/\sqrt{3}$. This is the familiar result of the vanishing of
$\sigma^{xy}$ when the Fermi surface consists of equal portions of
positive and negative curvature. One sees that the anisotropic small
angle contributions, which have a leading contribution proportional to
$\tau_M^3$, i.e. to
$1/T^3$, do not vanish there.

\section{Choice of Parameters and Comparison with Experiments}
We now describe the evaluation of the conductivities using experimental
data from ARPES and from the longitudinal transport in zero magnetic
field. The strategy is as follows. The anisotropies in the problem will
be determined from the anisotropy of the impurity scattering
$1/\tau_i(\theta) = 2\gamma(\hat k)/\hbar$ as determined from ARPES. For
simplicity, we shall let the density of states
$N(\theta)$ be responsible for the anisotropy of $1/\tau_i$, and hence
for the quantity $u(\theta)$, see  Eq.\ (21).  The measured
anisotropy of $1/\tau_i$ from the $\pi,\pi$ direction to the $\pi,0$
direction in the Brillouin zone then determines the velocity anisotropy
parameter $\rho$ from Eqs.\ (11,24). These parameterizations 
give the correct behavior at the edges of the Brillouin zone. The small
angle parameter
$\theta_c$ can then be found from the ``residual resistance" ratio ($RR$)
of the resistivity at
$T_c$ to the extrapolated value at
$T=0$. That is sufficient to determine the $T$-dependence of
$\cot\theta_H = \sigma^{xx}/\sigma^{xy}$. The magnitude of
$\cot\theta_H$ is then fixed by the effective mass $\hbar k_0/v_0$, which
we leave as an adjustable parameter. It will be seen that the effective
mass has a quite reasonable value.

The conductivities are given by Eqs.\ (25,33). The quantity $u$ which
appears in the differential equations for ${\bf M}$ and $Z$ is
defined as $u(\theta;T) = 
(\theta_c^2/24)\tau_M/\tau_i(\theta)
= (\tau_M/\hbar) (\theta_c^3/24)N(\theta)U(\theta,\theta)$. The measured
anisotropy \cite{pnas} of $\gamma(\hat k) = \hbar/2\tau_i$ is about 4.5.
We assign this to
$N(\theta)$, thus determining $\rho$ from Eq	.\ (24) as $\rho = 0.54$.
Therefore, we have
\begin{eqnarray}
1/\tau_i(\theta) &=& (1/\tau_0)(1-\rho\cos4\theta)^{-5/4},\nonumber
\\
u(\theta;T) &=& u_0(T)(1-\rho\cos4\theta)^{-5/4},\nonumber
\\
\rho &=& 0.54.
\end{eqnarray}
The quantity $u_0(T) = \tau_M(T)(\theta_c^2/24)/\tau_0$. $\tau_M(T)$ and $
\tau_0$ may be determined from ARPES as described in Ref.\ 3. From
ARPES, $\hbar/\tau_i \approx 0.24$ eV at $\theta=\pi/8$. So $\hbar/\tau_0
= 0.24$ eV. From the same source, $\hbar/\tau_M = 0.015(T/100)$ eV, where
$T$ is in Kelvin. This determines $u_0$ as $u_0(T)=
16(100/T)(\theta_c^2/24)$.  The remaining parameter $u_0$, that
is $\theta_c$, is determined from the resistance ratio $RR$ as follows.

The longitudinal resistivity in zero magnetic field $\rho^{xx}$ is given
by the inverse of $\sigma^{xx}$ from Eq.\ (25). Unfortunately we do not
have an analytic solution of the differential equation Eq.\ (19) for
$M_x$, although the coefficients are now known from Eq.\ (35). Numerical
integrations of Eqs.\ (19,25) are performed for different values of 
$u_0(T)= (1600/T)(\theta_c^2/24)$. For a given value of $\theta_c$,
$\rho^{xx}$ is almost precisely linear in $T$ above $T=100$ K. See Fig.\ 2. The
$T=0$ intercept of $\rho^{xx}$, $\rho_0$, is obtained by extrapolating the
found high temperature linear dependence. The $RR$ is defined as
$\rho^{xx}(T=100)/\rho_0$, and is typically about 8, say. This is obtained for
$u_0(T) \approx .022(100/T)$ so that the characteristic
small angle parameter  is indeed small, $\theta_c^2/24 \approx
0.0014$. 

\begin{figure}
\includegraphics[width=7cm]{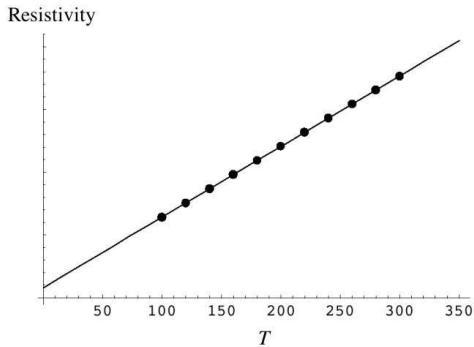}
\caption{Linear temperature fit to calculated resistivity (dots) from Eq.\ (33) for
$\theta_c^2/24= 0.0014$. The resistance ratio,
$\rho (100)/\rho (0)$, is 8.}
\label{fig2}
\end{figure}

All parameters are now fixed with the exception of $m^* = \hbar k_0/v_0$.
The numerical integrations of Eqs.\ (19,31,33) for $M_y,~Z$ and
$\sigma^{xy}$ respectively can be carried out for different values of
$T$ and combined with the previous result for $\sigma^{xx}(T)$ to
give  $\cot\theta_H(T)=\sigma^{xx}(T)/\sigma^{xy}(T)$:
\bq
\cot\theta_H(T)= \frac{0.82}{\omega_c\tau_M}~{m^*\over
m}~\frac{\int d\theta n(\theta) v_x(\theta) M_x(\theta)}{\int d\theta
n(\theta) v_x(\theta) Z(\theta)},
\eq
where we have used $\rho=0.54$, 
$\omega_c$ is the cyclotron frequency with the bare mass and $n(\theta)=
(1-0.54\cos 4\theta)^{-5/4}$. 

We now compare our result with the experiment of Chien, et al.\cite{ong}
They measured the Hall angle as a function of temperature for various
concentrations of Zn impurities in near-optimally doped YBCO. Their data
at $B=8$ T ($\omega_c = 1.28\times 10^{12}$ s$^{-1}$) is reproduced in
Fig.\ 3. 

\begin{figure}
\includegraphics[width=8cm]{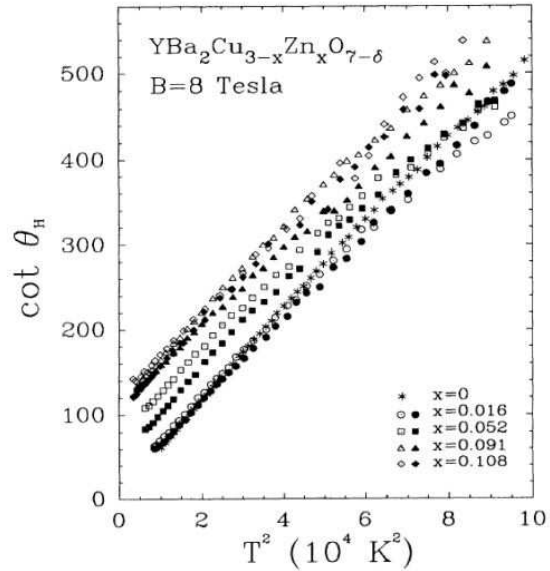}
\caption{Data of Chien, et al, Phys.\ Rev.\ Lett.\ {\bf 67}, 2088 (1991).
}
\label{fig3}
\end{figure}

We examine the data for zero Zn concentration. In Fig.\ 4, we have
replotted the $x=0$ points from Fig.\ 3 (large dots) and the
theoretical result (small dots) of Eq.\ (36) with   $m^*/m = 1.5$,
so adjusted to give the best agreement. It is important that $m^*$
turns out to be reasonable. We emphasize that the parameters we
picked were determined from ARPES experiments on BiSCO, while the
Hall data is on YBCO. Thus, our intent here is only to show that the
scattering rates characteristic of high-$T_c$ superconductors in the
normal state near optimal doping can by themselves account for the
temperature dependence of the Hall angle, not to demonstrate
quantitative agreement with a particular experiment. It is seen that
the experimental temperature dependence is fairly well reproduced.

\begin{figure}
\includegraphics[width=8cm]{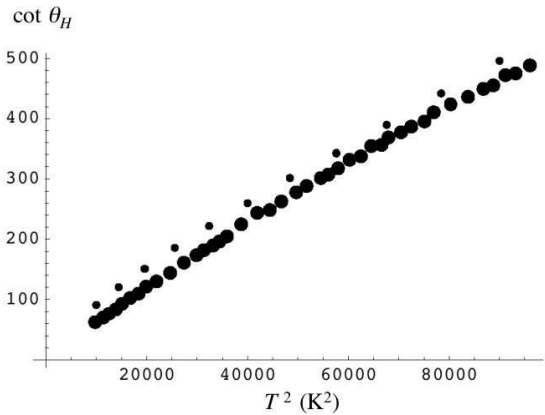}
\caption{Theory fit to $x=0$ data of Chien, et al.\cite{ong}}
\label{fig4}
\end{figure}

Notice that the experimental curves have a negative curvature on
a $T^2$ plot. While this might suggest a power law less than 2 and several
investigators have tried such a fit,\cite{ando,raffy} our picture
indicates that this behavior is a consequence of the fact that the conductivities
are each the sum of two terms with approximate $1/T^n$ and $1/T^{n+1}$
($\sigma^{xx}: n=1,~~~\sigma^{xy}: n=2$)
dependences.

The data of Ref.\ 6 and our theory both suggest that in the absence of
in-plane impurities such as Zn, the extrapolated $\cot\theta_H$ at
$T\rightarrow 0$ is zero. In-plane impurities will in general not give
scattering restricted to small angles. Indeed, their influence is evident
in the measured resistivity. We could include their effects by assuming them
to be isotropic scatterers which add a constant $1/\tau_z$ to the
isotropic linear in temperature MFL rate $1/\tau_M$. This has the effect
of an almost parallel upward shift of $\cot\theta_H$ so that it has a zero
temperature intercept proportional to $1/\tau_z$, i.e. to the in-plane
impurity concentration.
We illustrate this qualitatively using some data
from Ref.\ 9. In Fig.\ 5, we show the data (large dots) for
Bi$_2$Sr$_{2-x}$La$_x$CuO$_6$ with $x=0.44$ (optimal doping).
It is seen that the data does not extrapolate with a zero intercept.
We interpret this as signalling the presence of in-plane scatterers.
The theoretical curve (small dots) uses similar parameters as in
Fig.\ 4 but with an in-plane impurity contribution added to
$1/\tau_M$.

\begin{figure}
\includegraphics[width=7cm]{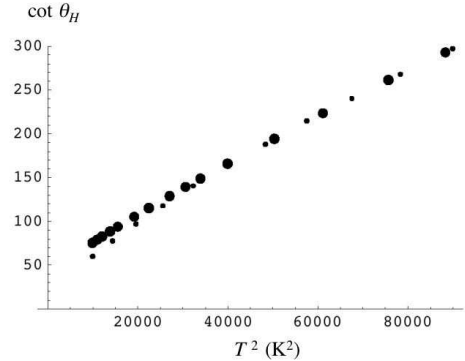}
\caption{Comparison of theory (small dots) including in-plane scatterers
with data (large dots) from Ref. 9, for optimally doped BSLCO.}
\label{fig5}
\end{figure}

\section{Complex Hall
conductivity} In this section, we comment on the recent
ac Hall effect results of Grayson, et al.\cite{matt} For low frequencies
and $T>T_c$, we can replace
$1\tau_M\approx 1/\tau_{tr}$ by $1/\tau_{tr} - i\omega$. This follows
immediately from Eqs.\ (3-5): the distribution function
$g({\bf{k}},\omega)$  and therefore the transport properties 
are obtained from the $\omega = 0$ results by this replacement.

From Fig.\ 2 it is seen that the resistivity is almost precisely linear in
$T$ in the normal state: $\rho^{xx} = a + bT$, where
$a$ is the (small) residual resistivity due to the small-angle
impurity scattering. Thus, $a/b$ is about 10, say. We have also found that
$\sigma^{xy}(T)$ is, in the temperature range of interest, very
closely of the form $c/T^2 +d/T^3$, with
$c/d\approx 0.01\,$K$^{-1}$. Therefore, neglecting quantities of
order 0.1, our prediction for the Hall angle is
\begin{eqnarray}
\theta_H(\omega, T) &\approx& \tan\theta_H =
\sigma^{xy}\rho^{xx} \nonumber \\
&\approx& A\left[ {1\over (1/\tau_{tr} -
i\omega)^2} + {c\over d}{1\over (1/\tau_{tr} - i\omega)}\right],
\end{eqnarray}
where $A$ and $c/d$ are
constant in temperature.  The latter ratio is now in units of
seconds; for our example parameters it is about $2\times
10^{-14}\,$s. We emphasize that a single relaxation rate
$1/\tau_{tr}$ enters all our expressions. The constant $A$ in Eq.\
(37) corresponds to the quantity
$\omega_H\Omega_p$ of Ref.\ 15. In that reference, Fig.\
4 shows that $A$ is indeed temperature and frequency independent.
The $c/d$ term is the conventional term. In Ref.\ 15, it was shown
that by itself it cannot account for the experimental data. Just as
the dc data shows deviations from $T^2$ behavior for
$\cot\theta_H$, we expect that the ac data should be fit by a
combination of the two terms in Eq.\ (37). At higher
frequencies or temperatures, the conventional term should finally dominate.
The complex Hall angle measurements convincingly demonstrate that
just one ineleastic transport rate determines all the frequency and
temperature dependence of the transport properties of the high-$T_c$
superconductors in the marginal Fermi liquid region, i.e. in their
normal state near optimum doping.

When calculating the ac conductivity, the limits of
applicability of the Boltzmann equation should be kept in mind. The
results are only valid for $\omega \tau\ll 1$. In this regime the
conservation laws (in terms of bare particles) completely determine
transport and the Boltzmann equation deals with them properly.
In a microscopic theory, the effects of small-angle scattering
calculated here appear as corrections to the coupling of the
external magnetic field to the carriers.
At high enough frequency these vertex corrections must vanish.
The crossover frequency and the behavior of transport properties in the
intermediate regime can only be determined by a microscopic  calculation.
Experimental results at high frequencies do depart from the predictions
of  Eq.\ (37).

\section{Concluding Remarks}
Here we discuss sundry issues related to the theory and calculations 
presented in this paper. We emphasize that this work pertains only to the Hall effect in 
optimally or overdoped samples. For the latter, the
linear-in-temperature scattering rate $1/\tau_M$ should be replaced in
the theory by a rate with the $T$-dependence of the observed resistivity.
In underdoped cuprates, additional physical considerations appear to
determine the transport and equilibrium properties.
\subsection{Magnetoresistance}
Earlier,\cite{prl} we gave a plausibility argument that the observed 
magnetoresistance would follow quantitatively
from the same considerations which lead to the observed Hall effect. We
leave it as a
future exercise to calculate this quantity directly from the solution of
the Boltzmann
equation as an extension of the present theory.

\subsection{Small-angle scattering, Fermi surface geometry, bilayer
splitting \ldots}
 
We have modeled the Fermi velocity through Eq.\ (23), which
with $\rho\approx 0.54$ is in itself 
enough to give the observed angular variation of the elastic part of
the single-particle relaxation rate. As we have seen, this also gives
agreement  with the observed temperature dependence of the Hall angle.
This  value of $\rho$ is 10\% away from the value 
at which the customary Hall coefficient would be zero. This
seems to us reasonable, since in a model with nearest neighbor Cu-O
hopping alone, the
Hall effect is zero at half-filling and with O-O hopping included, the
density for 
zero Hall effect shifts substantially towards hole doping. The optimal
doping composition
lies in the range 15-20\% hole doping. In this connection, it is important
to remember that we are using parameters obtained from experiments on
BSCCO to fit Hall data on YBCO. Therefore the precise numbers are not too
meaningful; we only argue that we have shown that the small-angle
scattering scenario gives the qualitative features of the experiments. 

In our numerical analysis, we have chosen the simplest parameterization of 
the Fermi velocity to show the plausibility of the small-angle effect in
relation to the experiments. Our parameterization  should not be expected
to represent the Fermi surface adequately. For the actual Fermi surface
several terms in Eq.\ (22) for the Fermi velocity would have to be included
with coefficients smaller than $\rho$. We note however that the successive
terms give compensatingly larger coefficients since the higher harmonics
produce larger derivative terms in Eq.\ (36) for $\sigma_{xy}$.
For a truer Fermi surface the advantage of analytical calculations is
lost. These remarks are of academic interest at the present time since the
actual Fermi surface is not accurately known.

More relevant is the fact that the Fermi surface of Bi-2212, the
ARPES data for which we have used here, has two sheets coming from the 
bilayer splitting. The bilayer splitting varies as a function of
angle, being largest in the $\pi,0$ direction. This arises because of
the geometry of the interlayer orbitals.\cite{ok} This angular variation
is essentially the same as that of the observed anisotropic elastic
scattering. The fact that the bilayer splitting is not resolved in the
experimental data we have used indicates the presence of an elastic
scattering mechanism which couples the layers. We surmise that for
interlayer or intralayer scattering due to impurities between the planes,
the same orbitals are involved and therefore the same angular dependence
arises in scattering.  In a single-layer model as treated here, this
effect is parameterized through the choice of the angle-dependence of the
Fermi velocity as in Eq.\ (23).

\subsection{Comments on related work}
As already remarked, although the analytical
expressions derived in our earlier work\cite{prl} are  correct, we
erred, as pointed out by V. Yakovenko, in our choice of the parameterization
in which they were evaluated. This error has been remedied here. Our more
complete analysis with a simpler parameterization in fact gives results
consistent with experiment. We have shown that not only do we get the right
temperature dependence but that with a reasonable effective mass, the
absolute value of the cotangent of the Hall angle is obtained within 10\% of
the data from the same parameterization.

In Ref.\ 19, R.\ Hlubina argues that the new term
discovered by us, if adequate to explain the Hall angle, would make an
unacceptable contribution to the longitudinal resistivity. Since our
parameters are actually determined by the measured
resistivity, this criticism is invalid. It is difficult to make a
direct comparison with the calculations of Ref.\ 19 since a quite
different parameterization of the Fermi surface is used there.
However, we can note that the parameterization family used in
Hlubina's evaluation does not include ours and gives a  much larger
conventional contribution to $\sigma^{xy}$ than the one we have used.

A recent paper of Carter and Schofield\cite{sch} addresses the main
point of  our original paper.\cite{prl} Their work consists of
two parts. One is analytical, the other is  a numerical solution of
equations we derived in Ref.\ 5.  We cannot comment on the
adequacy of the numerical work, but Carter and Schofield are indeed
correct that the customary term
(with a different temperature dependence than the new term) is zero
only for a particular choice of Fermi surface. This is of course
well-known: the traditional Hall angle is zero only for
``particle-hole symmetry" defined in terms of the curvature of the
Fermi surface.  However, as we have shown here, it is not necessary that
the customary term be zero (and all the temperature dependence come from
our new term) to get good agreement with experiment. It is sufficient
that the customary term is about a factor of four smaller than the new
term at
$T\approx 100K$. In fact, the experimental data for $\cot\theta_H$,
when plotted against $T^2$, usually shows a slight downward curvature.
This point has been discussed in Sec. I and is adequately demonstrated in
the experimental data shown in Fig3.\ 3-5.

The analytical calculation of Ref.\ 20, expressed in their Eqs.\ (7-10),
is not sufficiently general to contain the anisotropies required to show
the effect we have derived. Their calculation for $\sigma^{xy}$ is
equivalent to retaining only the impurity scattering
contribution to the conventional term, not the new contribution which we have
derived. See footnote 22 of Ref.\ 20 for a comment on this. Actually, the
analytical calculation of Ref.\ 20 is based on a circular Fermi surface
and we agree that such a parameterization never leads to a large enough
effect. 

\section{Acknowledgments}
We thank V. Yakovenko, D. Drew and M. Grayson for helpful
discussions of our analysis and A.J.\ Schofield for informative
communications. N.P.\ Ong and Y.\ Ando generously provided experimental
data. EA was supported in part by NSF grant DMR99-76665. The authors
acknowledge the hospitality of the Aspen Center for Physics where part of
this work was carried out.

\end{document}